# Development of the CMS Magnetic Field Map

Nicola Amapane [1,2] and Vyacheslav Klyukhin [3,4,*]

1. INFN Sezione di Torino, I-10125 Torino, Italy; nicola.amapane@cern.ch
2. Dipartimento di Fisica, Università di Torino, I-10125 Torino, Italy
3. Skobeltsyn Institute of Nuclear Physics, Lomonosov Moscow State University, RU-119992 Moscow, Russia
4. European Organization for Nuclear Research (CERN), CH-1211 Geneva, Switzerland
* Correspondence: vyacheslav.klyukhin@cern.ch

**Abstract:** This article focuses on pioneering work on the performance of the three-dimensional (3D) magnetic field map in the entire volume of the Compact Muon Solenoid (CMS) detector at the Large Hadron Collider at CERN. In the CMS heterogeneous magnetic system, the magnetic flux is created by a superconducting solenoid coil enclosed in a steel flux-return yoke. To describe the CMS magnetic flux distribution, a system of the primitive 3D volumes containing the values of the magnetic flux density measured inside the superconducting coil inner volume and modelled outside the coil across a special mesh of reference nodes was developed. This system, called the CMS magnetic field map, follows the geometric features of the yoke and allows the interpolation of the magnetic flux density between the nodes to obtain the magnetic field values at any spatial point inside a cylinder of 18 m in diameter and 48 m in length, where all the CMS sub-detectors are located. The geometry of the volumes is described inside one 30° azimuthal sector of the CMS magnet. To obtain the values of the magnetic flux density components across the entire azimuth angle of the CMS detector, rotational symmetry is applied.

**Keywords:** electromagnetic modelling; magnetic flux density; superconducting magnets; magnetic field map; magnetic flux; rotational symmetry; large-scale applications of superconductors

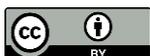

## 1. Introduction

The multi-purpose Compact Muon Solenoid (CMS) detector [1] at the Large Hadron Collider (LHC) [2] is designed and constructed to search for new particles and phenomena at the colliding beam centre-of-mass energy of 13.6 TeV. In particular, the production of the Higgs boson [3–5] with a mass of 125 GeV/$c^2$ was discovered by two LHC experiments ten years ago [6–8]. For the Higgs boson invariant mass reconstruction, the four-momenta of particles are used. This requires not only measuring the three-dimensional (3D) momenta of particles, but also their reliable identification.

In the CMS detector, precision silicon pixel and strip tracking sub-detectors [9,10], located in a strong magnetic field, measure momenta of the charged secondary particles generated in the collisions of the accelerated primary particle beams. In the magnetic field, the particle trajectories obtain a curvature [11], which depends on the magnetic flux density ***B***. A higher magnetic flux density provides a larger curvature and, as a result, a more accurate measurement of the charged particle momentum. In the CMS detector, the magnetic field is provided by a wide-aperture superconducting solenoid coil [12–14] enclosed into a 10,000-ton steel flux-return yoke [14,15]. The coil has an inner bore of 6 m in diameter, a length of 12.5 m, a thickness of 0.313 m, and is operated with the direct current of 18.164 kA, creating the central magnetic flux density |***B***$_0$| of 3.81 T.

The CMS magnetic system is of a heterogeneous type, i.e., the magnetic flux of 130 Wb created by the superconducting solenoid penetrates both non-magnetic and ferromagnetic materials of the experimental setup. The steel yoke of the magnet is used as a series of magnetized layers which are penetrated only by muons, making it possible to identify them and measure their momenta in a muon spectrometer [16–19]. In the azimuth angle, the yoke has dodecagonal shape and could be subdivided into 12 azimuthal 30° sectors labelled from S1 to S12 in positive direction of the azimuth angle $\varphi$ starting from the zero azimuth [1]. The reflection symmetry with respect to the longitudinal cross section of the magnet yoke is broken by the connecting brackets between the barrel wheel layers. The reflection symmetry with respect to the horizontal cross section of the yoke is broken by feet of the barrel wheels, and carts and keels of the endcap disks on both sides of the barrel wheels. The reflection symmetry with respect to the middle transverse plane of the magnet is broken by the absence of one coil turn that leads to a shift by several millimetres of the central value of the magnetic flux density with respect to the origin of the Cartesian coordinate system of the detector. Only rotational symmetry of the yoke azimuthal sectors is valid.

The large volume of the steel yoke around the coil and the inhomogeneity of the magnetic flux density in the yoke make direct measurements of the magnetic field inside the yoke blocks difficult. Existing methods for measuring the magnetic field in large steel blocks [20] allow only discrete measurements of the magnetic flux distribution in the magnet yoke. These discrete measurements are insufficient for the measurement of the muon momenta in a muon spectrometer. A computing modelling [21] of the magnetic system to obtain the distribution of the magnetic flux throughout the entire experimental setup achieves this goal.

The magnetic flux modelling is based on calculations of the magnetic field distribution performed with a 3D code TOSCA (two scalar potential method) [22], developed in 1979 [23] at the Rutherford Appleton Laboratory. To describe the solution of the magnetostatic problem, the TOSCA program relies on a database containing the geometrical description of the mesh used for the numerical finite element method [24] calculations. The mesh discretises the entire region of the model into quadrangular and triangular prisms. At each mesh node, the TOSCA database contains the computed values of the magnetic scalar potential [21] obtained as a result of the computation. The size of this database for the CMS magnet is of the order of 3.6 GB. It does not include the values of the magnetic flux density components, which are calculated from the database with a special post-processor module of the Opera-3d package [22]. The whole volume of the CMS magnet model is a cylinder of 100 m in diameter and 120 m in length, containing 8,759,730 nodes of the spatial finite element mesh.

The programs for simulation and reconstruction of the momenta of particles emerging from collision events require the knowledge of the value of the magnetic flux density components at the coordinates of space points along the trajectories of charged particles. To perform this task, only a part of the entire space of the 3D model is used. To describe the magnetic flux distribution in all the CMS sub-detectors, a cylinder of 18 m in diameter and 48 m in length is sufficient. The CMS field map is developed inside this cylinder and contains a total of 6,215,592 nodes storing the three components of the magnetic flux density extracted with the post-processor module from the TOSCA database in the form of 11,136 data tables. The magnetic flux density values at an arbitrary point in space is then obtained with high accuracy using a simple 3D linear interpolation over the eight adjacent nodes within the corresponding table. The organization of tables, called volumes, and the performance of the field map on the basis of these volumes are the subjects of this article.

The article is organized as follows: Section 2 describes the main principles of the CMS field map modelling and representation, Section 3 is devoted to the implementation of the *MagneticField* interface in the CMS software, and Section 4 presents the evolution of the magnetic field map. Finally, conclusions are drawn in Section 5.

## 2. Description of the Primitive Volumes Used in the Three-Dimensional CMS Magnetic Field Map

Preparing the magnetic field map in a heterogeneous magnetic system is not a trivial task. The sharp boundaries of the ferromagnetic parts in the CMS magnetic field map shown in Figure 1 would be practically impossible to be described using a regular mesh. Despite the scalar magnetic potential values being continuous at the steel–air interfaces, the total magnetic flux density is discontinuous following the difference in the permeability of the air and steel. To avoid the permeability uncertainty at the steel–air interface, a special solution is applied to the CMS field map preparation: the boundaries of the steel and air volumes of the magnetic field map are stepped by 0.1 mm out of the interface boundaries inside the adjacent magnetic system elements. This solution naturally leads to discretization of the entire field map space into primitive homogeneous air and steel volumes describing all the CMS magnet complicated geometry [21], for a total of 928 separate homogeneous volumes for each of the 12 azimuthal sectors of the CMS yoke.

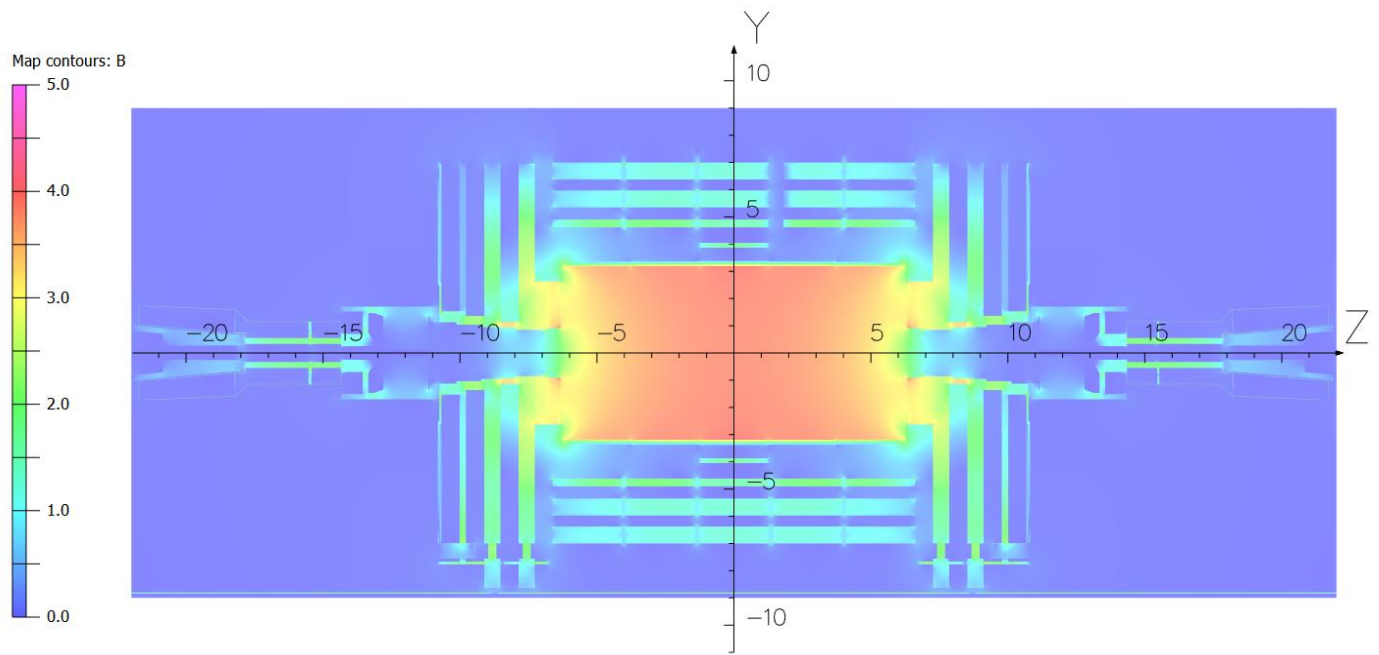

**Figure 1.** The longitudinal cross section of the CMS field map prepared with the 3D model of the CMS magnet [21] using the TOSCA program [22] for calculating the magnetic flux at an operating solenoid current of 18.164 kA. The color scale in Tesla describes the magnetic flux density distribution in an 18 × 44 $m^2$ region. The coordinate $Y$ and $Z$ axes show the distances from the coil centre in meters. The visible ferromagnetic materials of the magnet yoke include [21] five steel barrel wheels around the superconducting solenoid coil; two nose disks at each side of the coil; four steel endcap disks with the upper parts of their carts and keels on each side of the barrel wheels; steel absorbers and collars of the forward hadron calorimeter; ferromagnetic elements of the radiation shielding; and collimators of the proton beams. A 40 mm thick steel floor of the experimental cavern is visible as well.

The values of the magnetic flux density components in the nodes of these primitive volumes allow to obtain the component values between the nodes with a simple linear interpolation in the homogeneous material of the volume. To obtain the values of the components in the space between the boundaries of the volumes a linear extrapolation from the inner space of the closest volume is applied.

Figure 1 is prepared in the CMS detector $YZ$ plane. The origin of the CMS coordinate system is located in the centre of the superconducting solenoid coil, the $X$ axis lies in the LHC plane and is directed to the centre of the LHC machine, the $Y$ axis is directed upward and is perpendicular to the LHC plane, and the $Z$ axis makes up the right triplet with the $X$ and $Y$ axes and is directed along the vector of magnetic flux density **B** created on the axis of the superconducting coil [21].

To describe the geometries of the primitive volumes, only the azimuthal sector S1 shown in Figures 2 and 3 is used. This sector contains 928 primitive volumes in two semi sectors: one, *positive*, includes the volumes in the azimuth angle $\varphi$ from 0 to 15°, and another one, *negative*, is propagated in the azimuth angle $\varphi$ from –15° to 0. As seen in Figure 3, in the region of the tail catcher the volumes are displaced in the azimuth angle by 5° in the positive direction with respect to the ±15° angles. The geometries of 10,208 other primitive volumes in the azimuthal sectors S2–S12 are obtained by a symmetrical rotation of the volumes of sector S1 around the *Z* axis and applying the proper materials to the box volumes visible in Figure 3 in the region of the fourth layer of the muon drift tube chambers. The magnetic flux density values are calculated in each primitive volume separately in each azimuth sector S1–S12.

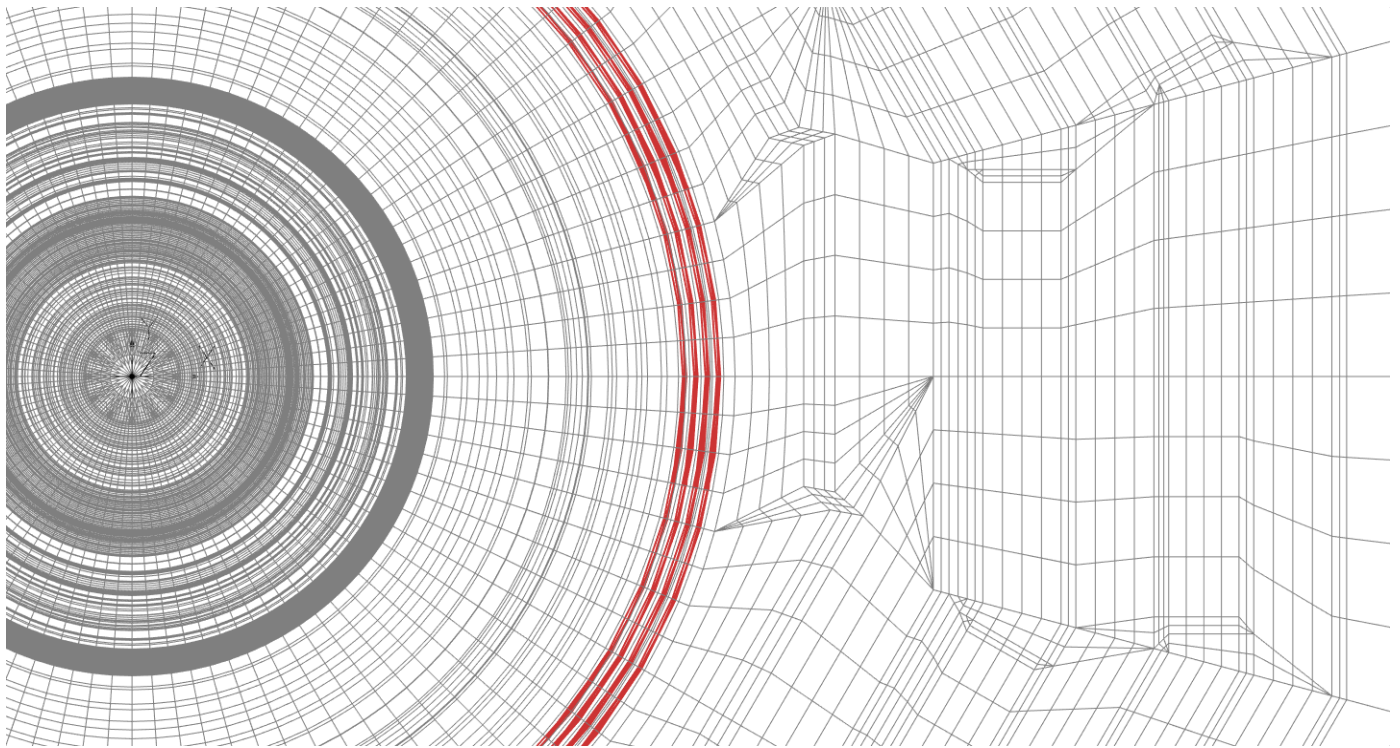

**Figure 2.** The finite element mesh of the CMS magnet model in the transverse section of the azimuthal sector S1 located at zero azimuth angle. The origin of the Cartesian *XYZ* coordinate system is shown on the yoke rotational axis. In a red color the modelled coil superconducting layers are displayed.

Within each volume the magnetic flux density components are calculated and tabulated for a set of nodes arranged on a grid over three to seven equidistant planes. From the software point of view, the primitive volume is an ASCII table including the nodes of all planes where each line includes the three coordinates of the node, the three components of the magnetic flux density ($B_x$, $B_y$, and $B_z$) in Tesla, the value of the relative permeability of the material in the primitive volume, and the value of the magnetic scalar potential in Ampere. According to the type of the node coordinates used in the primitive volumes, the volumes are of two types: $r\varphi z$-volumes (*r* or *R* are radii of the nodes from the coil axis in the transverse plane and $\varphi$ is the azimuth angle of the *RZ* plane), and *xyz*-volumes.

As shown in Figure 4, $r\varphi z$-volumes in the positive semi sector of the azimuthal sector S1 consist of five *RZ* planes, each spanning 3.75° in azimuth angle $\varphi$. These count 42.89% of the CMS magnetic field map volumes. To describe the barrel wheel blocks of the yoke with *RZ* planes is not possible because the connecting brackets between the barrel layers have a box shape and are orthogonal to the layers. Consequently, the *xyz*-volumes were introduced in this part of the yoke. These volumes in the positive semi sector of the azimuthal sector S1 are shown in Figure 5. The *xyz*-volumes consist of three to seven *YZ* planes and count the remaining 57.11% of the CMS magnetic field map volumes.

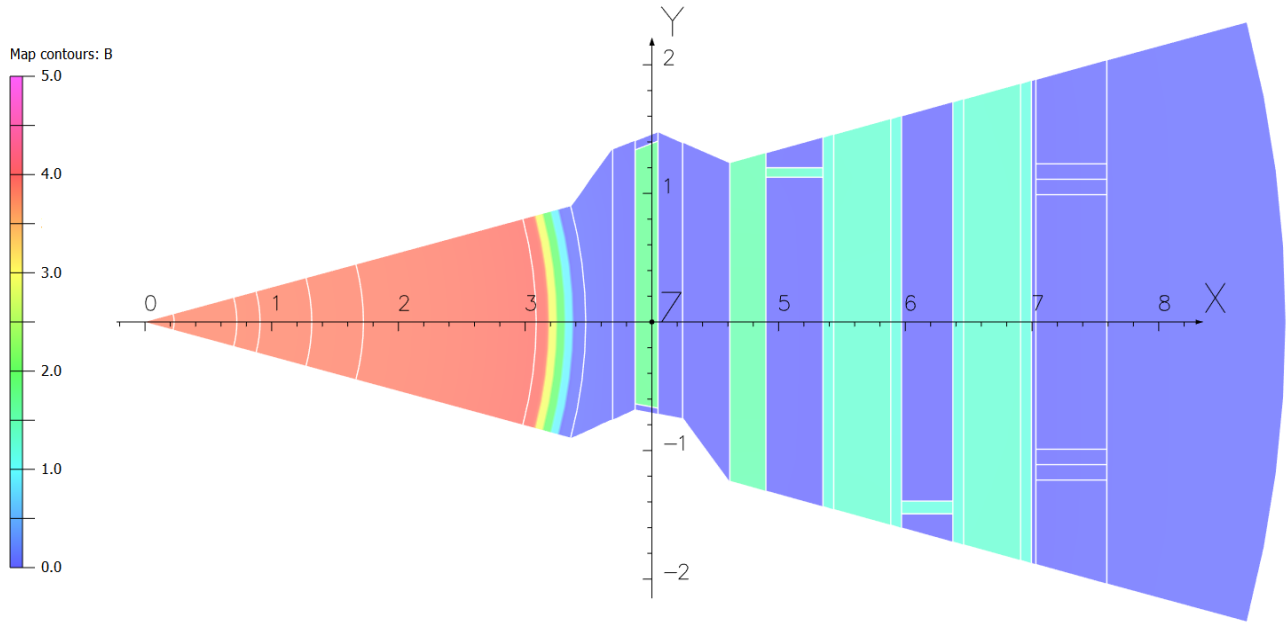

**Figure 3.** Simplified scheme of the primitive volumes in the CMS transverse section. The color scale in Tesla describes the magnetic flux density distribution inside the primitive volume elements. The coordinate *X* and *Y* axes show the distances from the coil centre in meters. The boundaries of the elements are represented with white lines. Four small box volumes at *X* = 7.03 to 7.59 m are used in the azimuthal sectors S9 and S11 to describe the barrel wheel feet in the region of 4th layer of the muon drift tube chambers. In these sectors, these volumes are modelled as steel instead of as air.

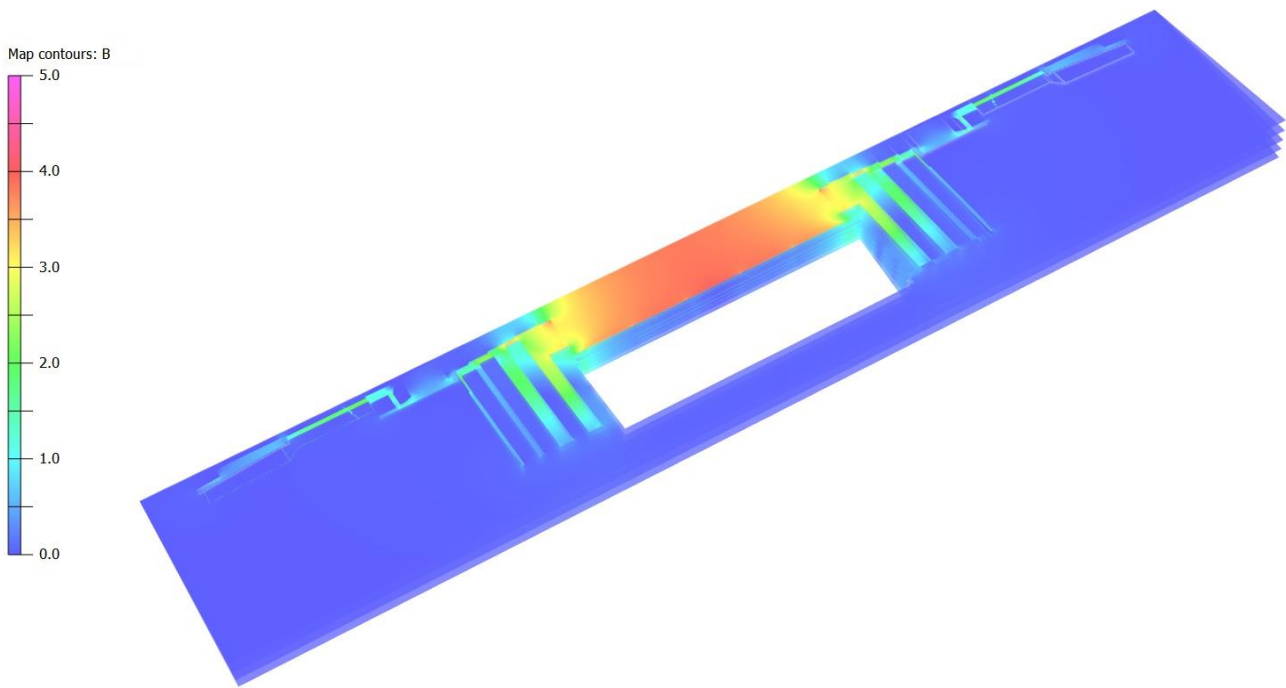

**Figure 4.** Magnetic flux density distribution in the *rφz*-volumes of the positive semi sector of the azimuthal sector S1. The color scale in Tesla has an increment of 0.5 T. The *rφz*-volumes occupy the whole 48 m length of the magnetic field map within a radius of 9 m, except of the shown empty space with a length of 13.24 m and a width of about 3.9 m that corresponds to the *xyz*-volumes presented in Figure 5.

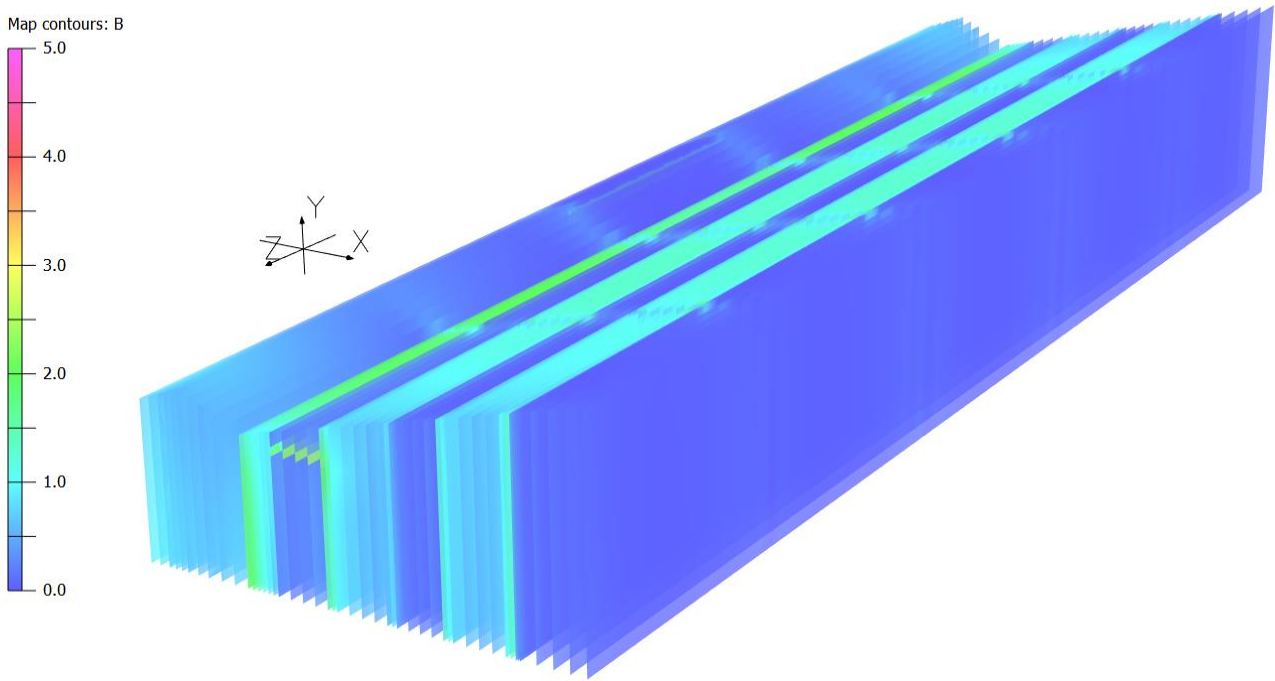

**Figure 5.** Magnetic flux density distribution in the *xyz*-volumes of the positive semi sector of the azimuthal sector S1. The color scale in Tesla has an increment of 0.5 T. Blue regions correspond to the air gaps between the barrel wheel steel layers, where the magnetic flux density is higher (green colours). Along the *Z* axis, the 155 mm air gaps at the sides of the central barrel wheel and the other two 125 mm wide air gaps between two external barrel wheels on each side of the central wheel are visible, as well as the connecting brackets between the first barrel wheel layer and the second one. The directions of the CMS Cartesian coordinate system are shown to help to understand the orientation and the dimensions of this part of the positive semi sector of the azimuthal sector S1 that propagates from $X$ = 3.690707 to 7.59 m in the limits of $Z$ = ±6.62 m within the azimuth angle $\varphi$ from 0 to 15°.

There are four types of primitive *rφz*-volumes: 113 *Tubs* (tube sector), 23 *TruncTubsOut* (tube sector truncated in the outer radius), 19 *TruncTubsIn* (tube sector truncated in the inner radius), and 44 *Cone* (conical sector) volumes in each semi sector. In Figure 3, the first seven primitive volumes are *Tubs* with outer radii of 0.23, 0.725, 0.908, 1.316, 1.724, 3.087345, and 3.478125 m. The first five radii are selected to have the mesh nodes at the radii, where the magnetic flux density was measured with a field mapping machine [25]. Between the last two, where the magnetic field drops stepwise from 3.94 to 0.07 T, the superconducting coil is located. The eighth primitive volume in Figure 3 is a *TruncTubsOut* and makes an interface between the *rφz*- and *xyz*-volumes at $X$ = 3.690707 m. The last *rφz*-volume in Figure 3 is a *TruncTubsIn* with the outer radius of 9 m. It makes an interface at $X$ = 7.59 m with previous *xyz*-volumes.

There are two types of primitive *xyz*-volumes: 109 *Box* and 156 *Trapezoid* volumes in each semi sector. In Figure 3, all the primitive volumes between $X$ = 3.690707 and 7.59 m are of those types. These volumes describe the steel yoke blocks, the air gaps, and the connecting brackets between the barrel wheel layers [21]. In Figure 5, the distribution of the magnetic flux density is shown in the *xyz*-volumes of the layers in the barrel wheel positive semi sector of the azimuthal sector S1.

Each semi sector within $|Z|$ < 24 m contains 24.35% volumes of *Tubs*-type, 4.96% volumes of *TruncTubsOut*-type, 4.09% volumes of *TruncTubsIn*-type, 9.48% volumes of *Cone*-type, 23.49% volumes of *Box*-type, and 33.62% volumes of *Trapezoid*-type.

The description of volume shape parameters is made in *XML* format and contains geometrical dimensions, material type, and position in the CMS coordinate system. Geometrical dimensions of the volumes are described in azimuthal sector S1 and the coordinates of the volumes in other azimuthal sectors are obtained from sector S1 with the rotational symmetry.

In Figure 6, the *XML* description of the *TruncTubsOut*-type *rφz*-volume is compared with initial human-friendly description made by hand. A special program is designed to prepare a conversion of the initial descriptions of all 928 volumes of the azimuthal sector S1 into the *XML* format.

```
Volume number = 1033 (YE/+1)
Type = TruncTubsOut
Phi1 (deg) = 0
Phi2 (deg) = 15
R_min (m) = 1.1349 (1.135)
R_max1 (m) = 6.9549 (6.955) at Phi=0
R_max2 (m) = 7.200242307 = 6.9549/cos(15) (7.20034=6.955/cos(15)) at Phi=15 deg
Z_min (m) = 7.2691 (7.269)
Z_max (m) = 7.8609 (7.861)
Material = ss400 steel
Nodes along R = 69 +(1)
Nodes along Phi = 4 +(1)
Nodes along Z = 4 +(1)
<Delta R, m> = 0.08435 at Phi=0; 0.0879 at Phi=15 deg
<Delta Phi, deg> = 3.75
<Delta Z, m> = 0.14795
```

```xml
<SolidSection>
  <TruncTubs name="V_1033" rMin="1.1349*m" rMax="7.20024*m"
    cutAtStart="6.9549*m" cutAtDelta="7.20024*m" cutInside="false"
    startPhi="0*deg" deltaPhi="15*deg" zHalf="0.2959*m"/>
</SolidSection>
<LogicalPartSection>
  <LogicalPart name="V_1033" category="unspecified">
    <rSolid name="V_1033"/>
    <rMaterial name="materials:Iron"/>
  </LogicalPart>
</LogicalPartSection>
<Algorithm name="global:DDAngular">
  <rParent name="cmsMagneticField:MAGF"/>
  <String name="ChildName" value="V_1033"/>
  <Numeric name="N" value="12"/>
  <Numeric name="StartCopyNo" value="1"/>
  <Numeric name="IncrCopyNo" value="1"/>
  <Numeric name="StartAngle" value="0*deg"/>
  <Numeric name="RangeAngle" value="360*deg"/>
  <Numeric name="Radius" value="0*m"/>
  <Vector name="Center" type="numeric" nEntries="3"> 0, 0, 7.565*m </Vector>
  <Vector name="RotateSolid" type="numeric" nEntries="3"> 0, 0, 0*deg </Vector>
</Algorithm>
```

**Figure 6.** A comparison of the *XML* format (on the right side) of the *TruncTubsOut*-type *rφz*-volume in the positive semi sector of the azimuthal sector S1 with the initial human-friendly description made by hand (on the left side). The cells of the magnetic field map used in this volume are 84.35 to 87.9 mm along the radius and 147.95 mm along the coil axis. The volume represents a semi sector of the first endcap disk of the yoke. An asterisk sign is used to separate a value and a value unit in an attribute or a child element.

The volume tables containing the magnetic flux density in the volume nodes are prepared with the Opera-3d post-processor module [22] and then are converted into a binary form to be used in the *MagneticField* interface of the CMS software to return the values of the magnetic flux density components for the coordinates of each requested space point along the trajectories of charged particles in the programs for simulation and reconstruction of the primary particle collision events. The pre-processor command files to prepare the data tables are created for azimuthal sector S1 and can be applied to other sectors using the central azimuth angle $\varphi$ of any other sector as a parameter.

## 3. *MagneticField* Interface in the CMS Software

The magnetic field map is typically accessed several hundreds of times during the simulation or the reconstruction of each single charged particle track; a fast and efficient implementation of the software interface to the field map is therefore essential to achieve an acceptable processing time. A dedicated software interface has been implemented in the CMS software for this purpose. It uses as inputs the geometric description of the primitive volumes and the data tables with magnetic flux density values within each volume, obtained as described in the previous section. Using this data, the problem of determining the components of the magnetic flux density at an arbitrary point in space is factorised in two steps: fast, optimized search of the volume containing the desired point, and simple interpolation of the values within the nodes defined for each volume. In addition, a parametrization of the magnetic flux density is used in the central solenoid region.

### 3.1. Volume Search

The algorithmic complexity of searching the volume containing a given point in space can be reduced exploiting the specific structure of the CMS geometry. The subdivision of volumes in the CMS magnetic field map is such that volumes can be organized in a hierarchical structure, with subsequent splitting of volumes by *R*, *φ*, and *Z*. At the top level, barrel and endcap regions are defined by boundaries in *Z*. Endcaps volumes are then organized in azimuthal sectors in *φ*, with each sector subdivided in layers along *Z*. Barrel volumes are organized in layers with boundaries defined by *R* (either the radius of cylindrical surfaces, or the tangent radius of the volume's innermost plane); each layer is subdivided in azimuthal sectors, and each sector in several rods of adjacent volumes. With volumes organized in such a hierarchical structure, volume finding is reduced to a simple one-dimensional binning problem (i.e., search in a sorted array of bounds) for each level of the hierarchy.

Moreover, the actual access patterns of particle simulation and reconstruction are very localized: the trajectory of a charged particle is followed along many steps, resulting in several consecutive queries within the same volume. Once the initial volume has been found, a volume caching mechanism is employed to skip further global volume searches until the track reaches the next volume. Cache hit rates of about 98% have been observed in typical applications [26], resulting in a substantial reduction of the required CPU time.

*3.2. Magnetic Flux Density Interpolation and Extrapolation Procedure*

Once the volume containing the desired spatial point has been found, interpolation within the nodes defined for that volume is used to obtain the result. The nodes are defined on a grid that is adapted to the shape of each volume. In the CMS magnet model, all rotation volumes (sectors of cylinders, cones, tubes, and truncated tubes) are centred along the *Z* axis; the barrel wheel elements are described by prisms and parallelepipeds. This arrangement of the volumes allows to cut the rotation volumes in each azimuthal sector of the CMS magnet by *RZ* planes with a constant step along the azimuthal angle, and to cut the prisms and parallelepipeds by planes parallel to the outer faces of the barrel wheels with a constant step between them. Each cutting plane is divided into cells with nodes where the magnetic flux density is calculated or measured. This definition of the grid of nodes within the primitive volumes allows a fast search for the needed nodes around the coordinates of the required point.

This idea comes from earlier works [27,28]. In [27], the preparation of the magnetic field map using the section of the magnetic system by *RZ* planes was considered. The magnetic flux density values were computed on the patches with a regular grid in each plane and any point in space was projected to the adjacent *RZ* planes. Then, on the plane, the radial ($B_r$) and axial ($B_z$) components of the magnetic flux density were computed by interpolation over nearest four nodes as shown in Figure 7 and expressed by Equation (1):

$$B_i = \frac{(B_{i1} \cdot r_2 + B_{i2} \cdot r_1)z_2 + (B_{i3} \cdot r_2 + B_{i4} \cdot r_1)z_1}{(r_1 + r_2)(z_1 + z_2)} \text{ for } i = r, z. \quad (1)$$

Similar expressions can be obtained in the *YZ* planes of sector S1 (or corresponded planes of the other azimuthal sectors) as well. The interpolation between adjacent planes is performed using either the azimuthal angle in the case of the *RZ* planes or a distance between the *YZ* planes.

Linear interpolation is a usual technique used in finite element analysis. The number of nodes in the magnetic field map volumes follows the number of the finite element nodes used in the CMS magnet model [21], where inside the superconducting coil the average length of the finite element is 65.5 mm in the radial direction and 86.8 mm in the axial direction. With the magnetic field obtained in the framework of the CMS magnetic field map, the degradation of the momentum resolution at the transverse momentum of 100 GeV/*c* is of –0.05 to +1.31% in comparison with the momentum resolution obtained with the constant homogeneous magnetic field [29].

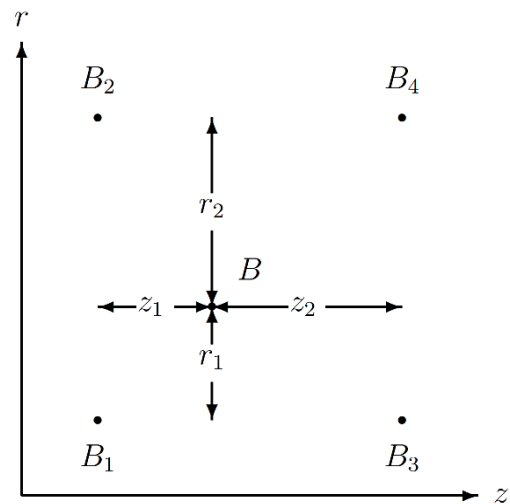

**Figure 7.** The scheme of the magnetic flux density interpolation in the *RZ* plane by four nodes close to the requested point.

Inside the superconducting solenoid, the CMS magnet model reproduces the measured field with an accuracy of 0.1% [20,25]. In the yoke steel blocks, the comparisons averaged over a set of seven measurements give the following differences between the modelled and measured values of the magnetic flux densities: 4.1 ± 7.0% in the barrel wheels and −0.6 ± 2.7% in the endcap disks [20].

*3.3. Parametrizations in the Central Region*

Most of the field map accesses during track reconstruction occur in the central region within the solenoid coil, where the CMS silicon tracker is located ($R < 1.15$ m, $|Z| < 2.8$ m). The field in this region follows closely the ideal solenoid behaviour [29]; in addition, this is the region where the magnetic flux density was measured in the field mapping campaign [25] mentioned above. A parametrization of the components of the magnetic flux density is used in this region, providing a result with good properties in terms of accuracy, smoothness, and CPU time efficiency. Different parametrizations are used depending on the required degree of accuracy and speed. For most applications, a fast parametrization of the $B_r$ and $B_z$ components expressed as a series expansion of the ideal finite solenoid formula with parameters extracted from a fit to the calculated values is used [26]. A simpler and faster polynomial parametrization of the $B_z$ component alone as a function of $Z$ and $R$ is used for specific applications where accuracy is not necessary and can be traded for execution speed. Finally, a parametrization of the measurements collected in the field mapping campaign using an expansion of the magnetic scalar potential over spherical harmonics represented in a cylindrical coordinate system [30] is available for applications that require the highest possible accuracy and for studies of systematic uncertainties related to the magnetic field.

**4. Magnetic Field Map Evolution**

The first CMS magnet model version used to create a space map of the magnetic flux distribution in the entire volume of the CMS detector comprised only 822,492 spatial nodes and was labelled 85_030919 [21]. Here, 8.5 represents the version of the Opera-3d package [22], and 2003.09.19 is the date of the CMS magnetic field map release. To describe the magnetic field map calculated with this model version, only 271 primitive volumes in a half of a single azimuth sector were used, with a total of 727,156 nodes storing the three components of the magnetic flux density. Using the rotational symmetry of the CMS magnet yoke and the reflections of the components with respect to the middle transverse plane, the magnetic field in all other sectors was determined by rotating the coordinates of the desired point to this sector and then by rotating back the components of the found magnetic flux density to the desired space point.

In 2007, the updated CMS magnet model version 1103_071212, containing 1,922,958 spatial mesh nodes in a half-cylinder containing the positive-*X* region of CMS yoke, was used to create a new CMS magnetic field map with 312 primitive volumes in the azimuthal sector S1. In comparison with the previous version, the new map included a separate description of the azimuthal sectors S3 and S4, which differ from the others because of the presence of chimneys for cryogenic and electrical leads [21], as well as the sector S11 with the barrel wheel feet. The field map in the sector S9 at the negative-*X* region was obtained by reflection of the magnetic field density values of sector S11 with respect to the vertical *YZ* plane. The magnetic flux density in the remaining sectors was determined based on the primitive volumes of the sector S1. These choices allowed the main specific features of special sectors to be modelled in an effective way, within the computing limits of the time. This map contained a total of 1,600,044 nodes storing the three components of the magnetic flux density.

At the end of 2008, the CMS detector was assembled in the underground experimental cavern and commissioned with cosmic muons. Based on the observed muon curvature, it was noted [30] that the magnitude of the magnetic field in the layers of the barrel wheels, especially the outermost ones, was overestimated in the CMS magnet model by several percent. This effect was understood to be caused by the compression of the returning magnetic flux of the solenoid within the insufficient extension of the model, which had the maximum outer radius of 13 m.

To minimize this effect, the half cylinder used for finite element calculations was expanded by increasing the outer radius from 13 to 30 m and its length from 40 to 70 m. Consequently, the number of spatial mesh nodes in the next model, 1103_090322, was increased to 1,993,452. The model included also the 40 mm thick steel floor of the underground experimental cavern [21]. The spatial extension in the model was necessary to properly describe the returning flux outside the yoke but was not necessary for the CMS map; primitive volumes did not need, therefore, to be extended.

In the following field map version, 14_120812, the whole CMS yoke was modelled in two separately calculated halves. This version included the enlarged fourth endcap disks which were newly installed in CMS, the endcap disk keels in the azimuthal sector S10, the outer parts of the radiation shielding with cylindrical gaps between the shield, and the collar of the forward hadron calorimeter. A total of 720 primitive volumes were used in each azimuthal sector, subdivided into two semi sectors of 15° azimuth angle each. A total of 8640 primitive volumes were extracted from the CMS magnet model in a sub-region with a diameter of 18 m and a length of 40 m.

In a further improvement, version 16_130503, the entire CMS magnet yoke was described in a single CMS magnet model with 7,111,713 nodes of the spatial mesh. The magnetic field map was described with 9648 primitive volumes due to a more detailed description of the regions outside the barrel wheels, where the last layer of the muon drift tube chambers is located. This map contained a total of 5,385,816 nodes storing the three components of the magnetic flux density.

Finally, the latest two versions, 18_160812 and 18_170812 of the CMS magnet model, differ from each other only by a slight refinement of the magnetization curve of one of three different kinds of steel used in the CMS magnet yoke. To complete the geometrical description of the magnetic system, the most distant part of the radiation shielding was added to both model versions at distances of ±21.89 m on both sides from the centre of the solenoid, as shown in Figure 1. Thus, the number of primitive volumes in the magnetic field map in all 12 azimuthal sectors was increased to 11,136, in a modelled sub-region with a length increased to 48 m. This final map contains a total of 6,215,592 nodes storing the three components of the magnetic flux density.

A recent study [29], based on the method of the magnetic field double integrals [31,32] that singles out the effect of the quality of the magnetic field on the momentum resolution of the charged particles, shows that the present approach for modelling the CMS magnetic field map works well in the presence of slightly magnetized materials inside the inner volume of the CMS superconducting solenoid.

## 5. Conclusions

The history of the CMS magnetic field map counts about 20 years. The pioneering approach applied to magnetic field map creation has been allowing the precise measurement of the momenta of charged particle, in particular the momenta of electrons (positrons) and muons used in the reconstruction of the invariant mass of the Higgs boson, as well as for all other results published by the CMS Collaboration.

Seven versions of the CMS magnet field map were developed using measurements and finite-element modelling of the magnetic flux distribution. The main clients of the magnetic field, which are a Monte Carlo simulation of events and the propagation of track parameters and errors in the event reconstruction, were made aware of the magnetic field with a software interface based on a volume-based geometry model. Computing efficiency in terms of execution speed and memory footprint was a driving requirement in the design of this interface. Volumes are defined as matching the magnetic properties of materials, so that the field within each volume is continuous and can be obtained with interpolation over a regular grid of points. Volumes are organized in a hierarchical structure optimized for fast global searching, and caching techniques allow simulation and track extrapolation algorithms to minimize the number of global volume searches.

**Author Contributions:** Conceptualization, V.K.; methodology, V.K. and N.A.; software, N.A. and V.K.; validation, V.K. and N.A.; writing—original draft preparation, V.K. and N.A.; writing—review and editing, V.K. and N.A. All authors have read and agreed to the published version of the manuscript.

**Funding:** This research received no external funding.

**Institutional Review Board Statement:** Not applicable.

**Informed Consent Statement:** Not applicable.

**Data Availability Statement:** Not applicable.

**Acknowledgments:** The authors are very thankful to the CMS Collaboration for availability to express and realize the ideas of the volume-based magnetic field map performance. This publication is agreed with the CMS Publication Committee.

**Conflicts of Interest:** The author declares no conflict of interest.